# AN OPTIMAL ANGLE OF LAUNCHING A POINT MASS IN A MEDIUM WITH QUADRATIC DRAG FORCE

**P. S. Chudinov**

*Department of Theoretical Mechanics, Moscow Aviation Institute, 125993 Moscow, Russia, E-mail: choudin@k804.mainet.msk.su*

**Abstract**

A classic problem of the motion of a point mass (projectile) thrown at an angle to the horizon is reviewed. The air drag force is taken into account with the drag factor assumed to be constant. Analytic approach is used for investigation. The problem of finding an optimal angle of launching a point mass in a medium with quadratic drag force is considered. An equation for determining a value of this angle is obtained. After finding the optimal angle of launching, eight main parameters of the point mass motion are analytically determined. These parameters are used to construct analytically six main functional relationships of the problem. Simple analytic formulas are used to solve two problems of optimization aimed to maximize the flight range of a point mass and minimize the initial speed of the point mass for getting to the given point on the plane. The motion of a baseball is presented as an example.

## 1. Introduction

The problem of the motion of a point mass under a drag force is considered in innumerable works [1 - 17]. Most of them solve the problem numerically. Analytical approaches to the solution of the problem are insufficiently developed. Meanwhile, analytical solutions are compact and convenient for a direct use in applied problems and for a qualitative analysis. Comparatively simple approximate analytic formulas to study point motion with quadratic drag force were obtained within the framework of this approach in [11 - 13]. One of the most important aspects of the problem is the determination of the optimal angle of throwing which provides the maximum range of throw [14], [15]. This paper shows how the formulas [11 - 13] are used to find the optimal angle of throwing and other kinematical parameters of the motion. These formulas are used for solution of two problems of optimization too. The first problem is solved to find an optimum angle of throwing of a point mass to get the maximum range in case when the point of incidence is above or below the point of throwing. The second problem is solved to find the optimum angle of throwing which ensures minimal initial velocity of the point mass for getting to the given point on the plane. Simple analytic formulas were obtained for solution of the both problems. These formulas make it possible to carry out a complete qualitative analysis without using numeric integration of point mass motion differential equations. The problems are solved by graphing the appropriate function. The given examples make a comparison between analytically obtained characteristics of motion and those calculated numerically. Numeric values of parameters are obtained by integration of point motion equations by the standard fourth-order Runge-Kutta method.

## 2. Equations of Motion, Local Solution and Analytical Formulas for the Main Parameters

The problem of the motion of a point mass under a drag force with a number of conventional assumptions, in the case of the drag force proportional to the square of the velocity, $R = mgkV^2$, boils down to the solution of the differential system [2]:

$$\frac{dV}{dt} = -g\sin\theta - gkV^2, \quad \frac{d\theta}{dt} = -\frac{g\cos\theta}{V}, \quad \frac{dx}{dt} = V\cos\theta, \quad \frac{dy}{dt} = V\sin\theta. \quad (1)$$

Here $V$ is the speed of the point, $m$ is the mass of the point, $\theta$ is the trajectory slope angle to the horizontal, $g$ is the acceleration due to gravity, $x, y$ are the Cartesian coordinates of the point, $k = \frac{\rho_a c_d S}{2mg} = const$ is the proportionality factor, $\rho_a$ is the air density, $c_d$ is the drag factor for a sphere, and $S$ is the cross-section area of the object.

The well-known solution of Eqs. (1) consists of an explicit analytical dependence of the velocity on the slope angle of the trajectory and three quadratures [2]

$$V(\theta) = \frac{V_0 \cos\theta_0}{\cos\theta\sqrt{1 + kV_0^2 \cos^2\theta_0 (f(\theta_0) - f(\theta))}}, \quad f(\theta) = \frac{\sin\theta}{\cos^2\theta} + \ln\tan\left(\frac{\theta}{2} + \frac{\pi}{4}\right), \quad (2)$$

$$t = t_0 - \frac{1}{g}\int_{\theta_0}^{\theta} \frac{V}{\cos\theta}d\theta, \quad x = x_0 - \frac{1}{g}\int_{\theta_0}^{\theta} V^2 d\theta, \quad y = y_0 - \frac{1}{g}\int_{\theta_0}^{\theta} V^2 \tan\theta d\theta. \quad (3)$$

The integrals on the right-hand sides of (3) are not taken in finite form. Hence, to determine the variables $t$, $x$ and $y$ we must either integrate (1) numerically or evaluate the definite integrals (3).

Using the integration of quadratures (3) by parts for enough small interval $[\theta_0, \theta]$, the variables $t$, $x$ and $y$ can be written in the form [11]

$$t(\theta) = t_0 + \frac{2(V_0 \sin\theta_0 - V \sin\theta)}{g(2+\varepsilon)}, \quad x(\theta) = x_0 + \frac{V_0^2 \sin 2\theta_0 - V^2 \sin 2\theta}{2g(1+\varepsilon)},$$

$$y(\theta) = y_0 + \frac{V_0^2 \sin^2\theta_0 - V^2 \sin^2\theta}{g(2+\varepsilon)}, \quad \varepsilon = k\left(V_0^2 \sin\theta_0 + V^2 \sin\theta\right). \qquad (4)$$

The function $V(\theta)$ in (4) is defined by relation (2). These formulas have a local form. Equations (4) enable us to obtain comparatively simple approximate analytical formulas for the main parameters of motion of the point mass [12]. We will give a complete summary of the formulas for the maximum ascent height $H$, time of flight $T$, particle speed $V_a$ at apex, flight distance $L$, time of ascent $t_a$, trajectory horizontal coordinate at apex $x_a$, the angle of incidence $\theta_k$ and the final velocity $V_k$ (see Figure 1):

$$H = \frac{V_0^2 \sin^2\theta_0}{g\left(2 + kV_0^2 \sin\theta_0\right)}, \quad T = 2\sqrt{\frac{2H}{g}}, \quad V_a = \frac{V_0 \cos\theta_0}{\sqrt{1 + kV_0^2 \cos^2\theta_0 f(\theta_0)}}, \quad L = V_a T,$$

$$t_a = \frac{T - kHV_a}{2}, \quad x_a = \sqrt{LH \cdot \cot\theta_0}, \quad \theta_k = -\arctan\left[\frac{LH}{(L-x_a)^2}\right], \quad V_k = V(\theta_k). \qquad (5)$$

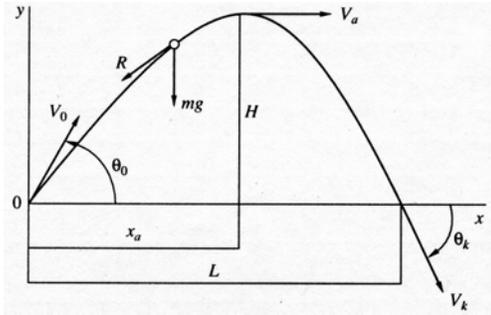

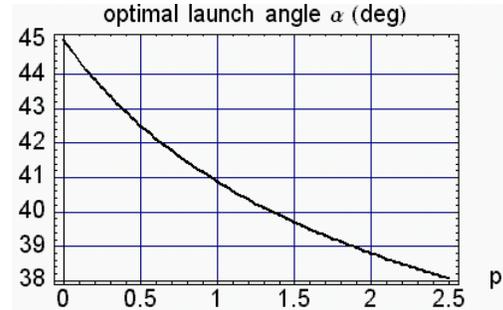

**FIGURE 1.** The main motion parameters.

**FIGURE 2.** The graph of the function $\alpha = \alpha(p)$.

In turn equations (5) make it possible to obtain simple analytic formulas for basic functional relationships of the problem $y(x)$, $y(t)$, $x(t)$, $x(\theta)$, $y(\theta)$, $t(\theta)$ [12]:

$$y(x) = \frac{Hx(L-x)}{x_a^2 + (L-2x_a)x}, \quad y(t) = \frac{Ht(T-t)}{t_a^2 + (T-2t_a) \cdot t}, \quad x(t) = \frac{w_1^2 + w_2 + w_1\sqrt{w_1^2 + \eta_1 w_2}}{\eta_2 w_1^2 + \eta_3 w_2},$$

$$x(\theta) = a\left(1 + \frac{1-n}{\sqrt{1 + b \cdot \tan\theta}}\right), \quad y(\theta) = c\left(d - \frac{2 + b \cdot \tan\theta}{\sqrt{1 + b \cdot \tan\theta}}\right),$$

$$t(\theta) = \frac{T}{2} + k_1 y(\theta) \mp \sqrt{(H - y(\theta))(k_2 - k_1^2 y(\theta))}. \tag{6}$$

The minus sign in front of the radical is taken on the interval $0 \leq \theta \leq \theta_0$ and the plus sign is taken on the interval $\theta_k \leq \theta \leq 0$. Here

$$n = L/x_a, \quad a = x_a/(2-n), \quad b = (L - 2x_a)/H, \quad c = H(n-1)/(2-n)^2,$$
$$d = 2 + H/c, \quad w_1(t) = t - t_a, \quad w_2(t) = 2t(T-t)/n, \quad \eta_1 = 2(n-1)/n,$$
$$\eta_2 = 2/L, \quad \eta_3 = 1/x_a, \quad k_1 = (t_a - 0.5T)/H, \quad k_2 = 0.25T^2/H.$$

The functions $x(\theta)$, $y(\theta)$, $t(\theta)$ are defined on the interval $\theta_k \leq \theta \leq \theta_0$. Thus, with the known motion parameters $H, L, T, x_a, t_a$, formulas (6) make it possible to construct functions $y(x), y(t), x(t), x(\theta), y(\theta), t(\theta)$.

## 3. The Finding of the Optimal Angle of Throwing

The formula for the range of throw is written as $L(\theta_0) = V_a(\theta_0) \cdot T(\theta_0)$. The optimal angle of throwing $\alpha$, which provides the maximum distance of flight, is a root of equation $dL(\theta_0)/d\theta_0 = 0$. Differentiating the $L(\theta_0)$ function with respect to $\theta_0$, after certain transformations, we obtain the equation for finding the angle $\alpha$ [13]:

$$\tan^2 \alpha + \frac{p \sin \alpha}{4 + 4p \sin \alpha} = \frac{1 + p\lambda}{1 + p(\sin \alpha + \lambda \cos^2 \alpha)}. \tag{7}$$

Here $p = kV_0^2$, $\lambda(\alpha) = \ln \tan\left(\frac{\alpha}{2} + \frac{\pi}{4}\right)$. The graph of function $\alpha = \alpha(p)$ calculated by means of (7) is submitted for Figure 2. The graph of this function is in full agreement with a numerically obtained graph the same function $\alpha = \alpha(p)$.

Figure 3 shows surface $\alpha = \alpha(k, V_0)$. This surface derived from the equation (7) under the conditions $0 \leq k \leq 0.001$ s²/m², $0 \leq V_0 \leq 50$ m/s. Figure 4 shows surface $L_{max} = L(\alpha(k, V_0))$. This surface derived from the equations (7) and $L = V_a T$.

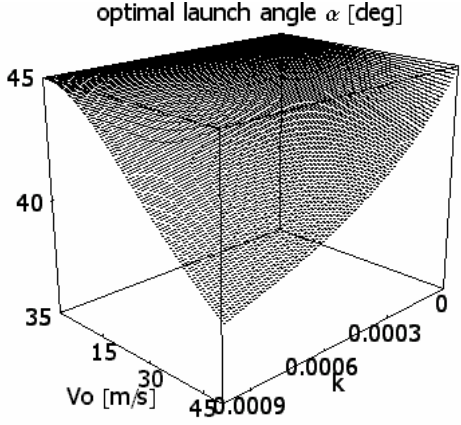

**FIGURE 3.** Surface $\alpha = \alpha(k, V_0)$.

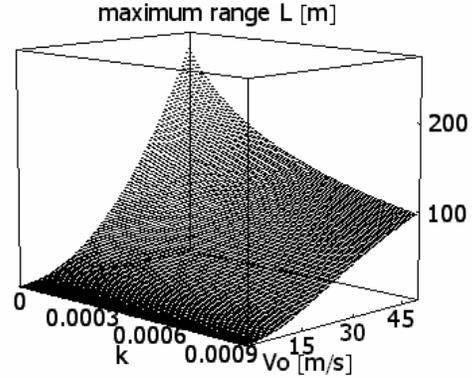

**FIGURE 4.** Surface $L_{max} = L(\alpha(k, V_0))$.

## 4. Determination of an Optimum Angle of Throwing in Case When the Point of Incidence is Above or Below the Point of Throwing

Let the point of incidence be on a horizontal straight line defined by the equation $y = y_1 = const$ (Figure 6). To solve the problem, let us make use of the point mass trajectory equation [12]

$$y(x) = \frac{Hx(L-x)}{x_a^2 + (L - 2x_a)x}.\qquad(8)$$

In this equation, the $H$, $L$, $x_a$ parameters are functions of initial conditions of throwing, $V_0, \theta_0$. They are defined by formulas (5). Let us substitute $y_1$ in the left-hand part of equation (8) and solve it for variable $x$. Thus, we shall get the flight range formula for the case under consideration

$$x(V_0, \theta_0, y_1) = \delta + \sqrt{\delta^2 - Ly_1 \cot\theta_0},\qquad(9)$$

where $\delta = \frac{L}{2} + \frac{y_1}{H}\left(x_a - \frac{L}{2}\right)$. For the given values of the $V_0, y_1$ parameters, range $x$ is the function of $\theta_0$, the angle of throwing. To find the optimum angle of throwing $\theta_0^*$ and the maximum range $x_{max}$ gained with the given $V_0, y_1$ values, it is sufficient to construct a graph of the function. Coordinates of the function maximum point will define $\theta_0^*$ and $x_{max}$ values. As an example, let us analyze the motion of a baseball with the drag factor $k = 0.000548$ s²/m² [10]. Other parameters of the motion are given by the following values: $g = 9.81$ m/s², $V_0 = 40$ m/s, $y_1 = \pm 20$ m. Computation results are presented in Figures 5 and 6. The graph of the $x(\theta_0)$ function for $y_1 = 20$ m is given in Figure 5. One can see from the graph that the maximum range $x_{max} = 84.4$ m is attained when the angle of throwing $\theta_0^* = 47.5°$. A numeric analysis of the problem based on integration of the motion equations (1), yields as follows: $\theta_0^* = 47.5°$, $x_{max} = 84.0$ m.

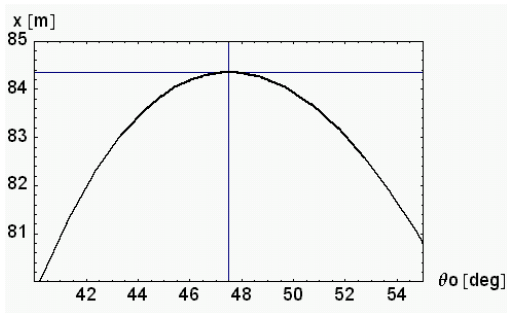

**FIGURE 5.** The graph of the $x = x(\theta_0)$ function.

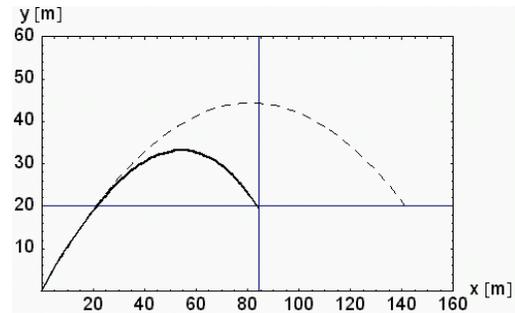

**FIGURE 6.** The graph of the $y = y(x)$ function.

The trajectory of the point mass is shown in Figure 6 as a solid line. The broken line presents the trajectory of the point mass launched under the same initial conditions $V_0 = 40$ m/s, $\theta_0^* = 47.5°$ with no air drag ($k = 0$).

## 5. Determination of the Optimum Angle of Throwing the Point Mass Which Ensures Minimized Initial Velocity for Getting to the Given Point on the Plane

Suppose that it is necessary to get from the origin of coordinates $O(0,0)$ to the given point $A(x_1, y_1)$ with the minimized initial velocity $V_0$ (Figure 8). Let's solve this problem using trajectory equation (8). We substitute coordinates $x_1, y_1$ of the given point A in the equation (8) and define the initial velocity $V_0$ from it. The resulting formula is as follows

$$V_0(\theta_0, x_1, y_1) = \frac{2}{\sqrt{\dfrac{a}{3d^2 + 2d} - 2k \sin\theta_0}}.\qquad(10)$$

Here $\quad a = k\left(\sin\theta_0 + 2\cos^2\theta_0 \ln\tan\left(\dfrac{\theta_0}{2} + \dfrac{\pi}{4}\right)\right),\qquad b = \dfrac{agx_1^2(1 + \tan^2\theta_0)}{4(y_1 - x_1 \tan\theta_0)},$

$$c = \frac{2by_1 \cot\theta_0 - x_1(b+2)}{x_1(b+6)},\qquad d = c + \sqrt{c^2 - \frac{b}{b+6}}.$$

As an example, let's calculate motion of a baseball with the following parameters

$$k = 0.00095 \text{ s}^2/\text{m}^2,\quad g = 9.81 \text{ m/s}^2,\quad x_1 = 60 \text{ m},\quad y_1 = 20 \text{ m}.\qquad(11)$$

The graph of the $V_0(\theta_0)$ function is presented in Figure 7. It follows from the graph that the minimal initial velocity $V_0^{min} = 38.3$ m/s is achieved for the throwing angle $\theta_0^* = 48.8°$. A numeric analysis gives $\theta_0^* = 48.8°$ and $V_0^{min} = 38.0$ m/s. An error of an analytic determination of the $\theta_0^*$, $V_0^{min}$ parameters in the given problem does not exceed 1%. The trajectory of the point mass for values (11) is shown in Figure 8 as a solid line. The broken line presents the trajectory of the point mass for the no drag case.

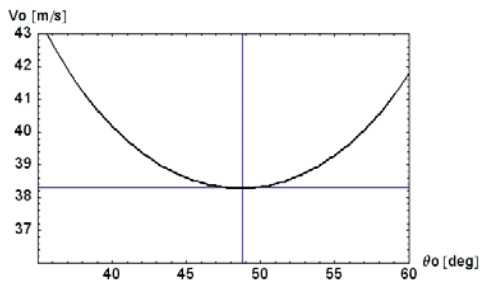

**FIGURE 7.** The graph of the $V_0 = V_0(\theta_0)$ function.

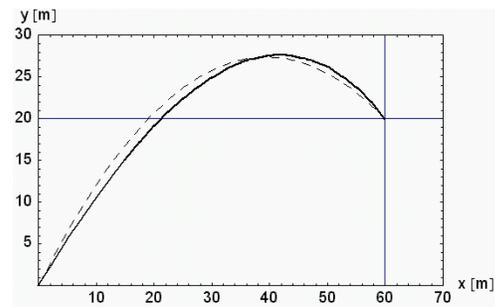

**FIGURE 8.** The graph of the $y = y(x)$ function.

## 6. Conclusion

The proposed approach based on the use of analytic formulas makes it possible to significantly simplify a qualitative analysis of the motion of a point mass with air drag taken into account. All basic parameters of motion, functional relationships and various problems of optimization are described by simple analytic formulas. Moreover, numeric values of the sought variables are determined with an acceptable accuracy. Thus, set formulas [11 – 13] and (5) – (10) make it possible to carry out complete analytic investigation of the motion of a point mass in a medium with drag the way it is done for the case of no drag. As a whole, the collection of formulas (4)–(10) makes it possible to widen considerably the possibilities of studying the problem of the point mass motion with quadratic drag force and to supplement numeric methods with analytic ones.

## References

1. L. Euler , "An investigation of the true curve described by a projectile in air or in any other medium" , In *Research in Ballistics,* Edited by Euler L., 455 – 494  ( Fizmatgiz, Moscow, 1961)
2. B.N. Okunev , *Ballistics*  (Voyenizdat,    Moscow, 1943)
3. N. de Mestre, *The Mathematics of Projectiles in Sport* (Cambridge University Press, New York, 1990)
4. R. K. Adair, *The Physics of Baseball*  (Harper&Row Publishers Inc.,New York, 1990)
5. D. Hart, T. Croft, *Modelling with Projectiles* ( Ellis Horwood Limited, West Sussex, 1988)
6. E.T. Whittaker, *A Treatise on the Analytical Dynamics of Particles and Rigid Bodies* (Cambridge University Press, London, 1961)
7. S. Timoshenko, D.H. Young, *Advanced Dynamics*  (McGraw-Hill, New York, 1948)
8. G.W. Parker, "Projectile motion with air resistance quadratic in the speed", *Am. J. Phys.* **45**, 606 – 610, 1977
9. H. Erlichson, "Maximum  projectile range with drag and lift, with  particular application to golf", *Am. J. Phys.* **51**, 357 - 362 , 1983
10. A. Tan , C.H. Frick , O. Castillo , "The fly ball trajectory: an older approach revisited", *Am. J. Phys.* **55**, 37 – 40, 1987
11. P.S. Chudinov , "The motion of a point    mass in a medium with a square law of drag", *J. Appl. Maths Mechs* **65**(3), 421 – 426, 2001
12. P.S. Chudinov ,  "The motion of a heavy particle in a medium with quadratic drag force", *International J. Nonlinear Sciences and Numerical Simulation* **3**(2), 121-129, 2002
13. P.S. Chudinov, "An optimal angle of launching a point mass in a medium with  quadratic drag force", *Indian  J. Phys.* **77B,** 465-468, 2003
14. C.W. Groetsch , "On the optimal angle of   projection in general media", *Am. J. Phys.* **65**, 797 – 799, 1997
15. R.H. Price, J.D. Romano , "Aim high and go far - Optimal projectile launch angles greater than 45° ", *Am. J. Phys.* **66**, 109 – 113, 1998
16. J. Lindemuth, " The effect of air resistance on falling balls", *Am. J. Phys.* **39**,  757 - 759, 1971
17. P. Gluck, "Air resistance on falling balls and balloons", *Phys. Teach.* **41**, 178-180, 2003